\documentclass[twoside,fleqn]{article}
\usepackage{espcrc2}
\usepackage{psfig}

\title{Study of stochastic estimates of quark loops with unbiased subtraction
\thanks{Talk presented by N. Mathur at Lattice 2002.}
\thanks{This work is supported in part by the CCS of 
University of Kentucky and U.S. DOE
        under grant numbers DE-FG05-84ER40154 and DE-FG02-02ER45967.}}

\author{Nilmani Mathur and Shao-Jing Dong \address{Department of Physics and Astronomy,
University of Kentucky, Lexington, KY 40506-0055, USA\\}}
\begin{document}
\begin{abstract}
Stochastic noise estimator method is a powerful tool to calculate the
disconnected insertion involving quark loops. We study the variance
reduction technique with unbiased subtraction. We use
the complex $Z_2$ noise to calculate the quark loops
on a $16^3 \times 24$ lattice with $\beta = 6.0$ and $\kappa$ = 0.154.
Unbiased subtraction method is performed by using hoping parameter expansion. 
We report on the variance reduction for the
point-split vector current as a function of the number of subtraction 
terms and the number of noise used.

\end{abstract}

\maketitle

\section{INTRODUCTION}
Disconnected quark loops originates from the vacuum polarization of 
sea quarks of the interacting operator.  
Whether we do lattice calculation for quenched or dynamical QCD, these
 quark loops contribute significantly 
to various hadronic matrix elements. For example, for nucleonic scalar 
($\pi NN$-$\sigma$ term) \cite {sc},
vector (strange magnetic moment) \cite {vec1,vec2}, flavor-singlet axial (quark spin) 
\cite {spin}, and tensor (quark orbital angular momentum) \cite{ang} channels, it is 
absolutely essential to consider disconnected quark loops.  In particular, the
strangeness content of 
the nucleon comes exclusively from these disconnected loops.

However, disconnected quark loops are difficult to simulate as they contain 
both diagonal and off-diagonal elements of the large inverse fermion matrix ($M$). 
Even for a moderate size lattice (e.g. $16^{3} \times 24$) one needs to invert 
a $10^{6}\times10^{6}$ matrix which requires an enormous amount of computation time. 
Instead, one can use noise method to estimate required 
traces \cite{noise1,noise2} to obtain disconnected loop contribution. 
This study focuses on the stochastic estimation of  
the strangeness magnetic form factor ($G_{M}^{s}(0)$) of 
the nucleon and reports a systematic analysis 
about the number of noises required to extract its signal.

Experimentally we are still uncertain about the value (even the sign) of 
the strangeness magnetic moment of the nucleon. 
SAMPLE \cite{sample} and HAPPEX \cite{hpx} expts. reported
$G_{M}^{s}(0) = 0.01 \pm 0.29 \pm 0.31 \pm 0.07$ and $G_{E}^{s} + 0.39 G_{M}^{s}
({\hbox{at}} \,Q^{2} = 0.477 GeV^{2}) = 0.025\pm 0.020 \pm 0.014$, 
respectively. Prediction from theoretical models vary  in a wide range 
($-0.75$ to $+0.30\mu_{N}$) \cite{st}. Lattice QCD results also differ in conclusion.
Our previous studies \cite{vec1} 
suggest $G_{M}^{s}(0) = -0.28 \pm 0.10$, while ref. \cite{vec2} reported very tiny signal for 
$G_{M}^{s}$. This work is aimed at studying why there is such a difference.

\section{Noise method and unbiased subtraction}
Disconnected quark loop calculation by stochastic noise method have been detailed in refs. 
\cite{noise1,sc,vec1,vec2,spin,ang}. They used random noises to estimate various 
traces involving fermion matrix. Using a set of random noise vectors $\eta$, one can estimate 
the trace of a $N \times N$ matrix $A$ as \cite{noise1,noise2,unbiased} 
\begin{eqnarray}
\hspace{0.5in}\hbox{Tr}\left( A \right) &\equiv& E \left[<\eta^{\dagger} A \eta>\right].
\end{eqnarray}  
Of course, this trace 
will be an approximation for finite number of noises 
and the variance of the estimator depends on the choice of the noise. 
It has been demonstrated that $Z_{2}$ noise \cite{noise1} is 
the optimal noise with minimum variance \cite{noise2}. 
For a given $L$ number of $Z_{2}$ noises, variance of this estimation is given by \cite{noise1,noise2,unbiased}
\begin{eqnarray}
\sigma^{2}_{A}&\equiv& {\hbox{Var}} \left[<\eta^{\dagger} A \eta>\right]
\,=\,{1\over L} \sum^{N}_{m \ne n} {{\biggl|} A_{mn}{\biggr|}}^{2}. 
\end{eqnarray}
This variance can further be reduced by the method of unbiased subtraction \cite{unbiased}, 
where a set of $P$ traceless matrices ($Q$) are subtracted from the matrix $A$ as 
\begin{eqnarray}
\hbox{Tr} (A)\,=\, E {\left[\left< \eta^{\dagger} \left(A 
- \sum^{P}_{p=1} \lambda_{p} Q^{(P)} \right) \eta \right> \right]},
\end{eqnarray}
where $\lambda's$ are some variational coefficients. 
Corresponding reduced variance will be \cite{unbiased} 
\begin{eqnarray}
\sigma^{2}_{A}(\lambda) 
&=&{1\over L} \sum_{m\ne n} \left|A_{m,n} 
- \sum^{P}_{p=1} \lambda_{p} Q^{(P)}_{m,n}\right|^{2}.
\end{eqnarray}
This subtraction is unbiased in the sense that it does not change the expectation value of 
$\hbox{Tr}(A)$.
Choice of these traceless subtraction matrices ($Q$) should be such that they match the 
off-diagonal behavior of the matrix $A$. For disconnected loop calculation $A$ will 
be replaced by the inverse fermion matrix $M^{-1}$. 
Previously it was shown that the above variance for $M^{-1}$ 
can be reduced substantially \cite{unbiased,vec1,vec2,ang} by using a set of 
traceless matrices 
obtained from the hoping parameter expansion of the  fermion matrix M as 
\begin{eqnarray}
M^{-1} = I + \kappa D + \kappa^{2} D^{2} + \kappa^{3} D^{3} +  \kappa^{4} D^{4} + \cdots  
\end{eqnarray}
For the point split conserved current, disconnected quark loop can be written as
\begin{eqnarray}
{\hbox{Loop}} = &&\hspace*{-0.2in}\sum_{\vec{x}}{e^{-i\vec{q}\cdot\vec{x}}}{\hbox{Tr}}
{\biggr[}M^{-1}(x,x+\mu)(1+\gamma_{\mu})
U^{\dagger}_{\mu}(x)\nonumber\\
&&\hspace*{0.05in} -\,\, M^{-1}(x+\mu,x)(1-\gamma_{\mu})U_{\mu}(x)
{\biggr]},
\end{eqnarray}
Before subtracting each matrix ($I$, $\kappa D$, $\kappa^{2}D^{2}$ etc. of Eq.(5)) 
from $M^{-1}$ in Eq.(6), one should 
make sure that it does not change the loop expectation value. 
In fact, all matrices with $M^{-1}$ substituted with even order of $D$ are 
traceless in Eq.(6). First and second terms ($I$ and $\kappa D$) are 
also traceless. However, starting from $\kappa^{3} D^{3}$, all odd orders in $D$ are 
not traceless. So, to subtract an odd order term another matrix is need to be subtracted form it 
so that the resulting matrix is traceless. For example, for $\kappa^{3} D^{3}$ term, one needs to 
subtract following plaquette terms from the loop:
\begin{eqnarray*}
%
&&\hspace*{-0.2in}-\,8\kappa^{3}\sum_{x} e^{i\vec{q}.\vec{x}}
\,\,\sum_{\nu}{\hbox{Tr}}\, \fbox{}_{\mu\nu}(x)
\,+\, {\hbox{Tr}}\, \fbox{}^{*}_{\mu\nu}(x-\nu),\nonumber\\
%
&&\hspace*{-0.2in}-\,8\kappa^{3}\sum_{x} e^{i\vec{q}.\vec{x}}
\,\,\sum_{\nu}{\hbox{Tr}}\, \fbox{}^{*}_{\mu\nu}(x)
\,+\, {\hbox{Tr}}\, \fbox{}^{}_{\mu\nu}(x-\nu),
\end{eqnarray*}
corresponding to current in the $(1+\gamma_{\mu})$ and $(1-\gamma_{\mu})$ 
respectively. 
One should notice that in each term plaquettes are at position $x$ and $x - \nu$, 
and so, at $|\vec{q}| \ne 0$ one cannot use translational invariance due to 
the Fourier transformation factor. Similarly, one needs to 
subtract some chair diagrams from $\kappa ^{5} D^{5}$ terms to 
make it traceless.
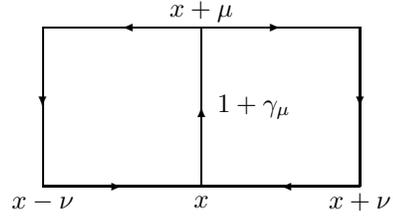
\begin{figure}[h] 
\vspace*{-0.6in}
\begin{picture}(140,140)(-70,0)
\put(40,40){\vector(0,1){30}}
\put(40,70){\line(0,1){30}}
\put(40,100){\vector(-1,0){30}}
\put(10,100){\line(-1,0){30}}
\put(-20,100){\vector(0,-1){30}}
\put(-20,70){\line(0,-1){30}}
\put(-20,40){\vector(1,0){30}}
\put(10,40){\line(1,0){30}}
\put(40,100){\vector(1,0){30}}
\put(70,100){\line(1,0){30}}
\put(100,100){\vector(0,-1){30}}
\put(100,70){\line(0,-1){30}}
\put(100,40){\vector(-1,0){30}}
\put(70,40){\line(-1,0){30}}
\put(40,34){\makebox(0,0){$x$}}
\put(100,34){\makebox(0,0){$x+\nu$}}
\put(-20,34){\makebox(0,0){$x-\nu$}}
\put(60,70){\makebox(0,0){$1+\gamma_{\mu}$}}
\put(40,106){\makebox(0,0){$x+\mu$}}
\end{picture}
\vspace*{-0.65in}
\caption{Plaquette term associated with current corresponding to $1+\gamma_{\mu}$. This needs to 
be subtracted from $\kappa^{3}D^{3}$ term to make it traceless.}
\vspace*{-0.45in}
\end{figure}

\section{Results}
Numerical simulation was done on a $16^{3}\times 24$ lattice at $\kappa = 0.154$
with 60 configurations where each configuration is separated by 20,000 sweeps.
Unbiased subtraction is done with terms up to $\kappa^{4} D^{4}$.  
We systematically study the signal for the strange quark form factor as a function of
 the number of 
complex $Z_{2}$ noises used per configuration. In Fig. 2 we plot    the summed ratio of 
three to two point functions as a function of the time slice,  
from which one can obtain the magnetic form factor (see ref \cite{vec1} for notations). 
Valence and sea quark mass is kept fixed at 
$\kappa = 0.154$.
First sub-figure is with 300 noises without any unbiased subtraction. Next 4 sub-figures are 
results with unbiased subtraction with different number of noises (30, 100, 200 and 300, 
respectively). It is clear from these sub-figures that we do not find any signal up to 200 noises 
and the signal becomes prominent at around 300 noises. The fitted slopes for 
100, 200, and 300 noises are 
$-0.052 \pm 0.09$, $-0.060 \pm 0.048$ and $-0.092 \pm 0.040$, respectively. This implies that 
the signal can only be extracted 
out at around 300 $Z_{2}$ noises. Slope for the 300 noise case agrees well 
to our previous calculation where we used subtraction terms up to
$\kappa^{2} D^{2}$. 
Since this result and previous result agree at one $\kappa$, 
we do not carry out calculation for other $\kappa$ values. 
In our previous calculation we obtained $G^{s}_{M}(0) = -0.28\pm0.10$ and this 
systemic study of noise versus signal supports that result.
\begin{figure}[t]
\vspace*{-0.35in}
\includegraphics{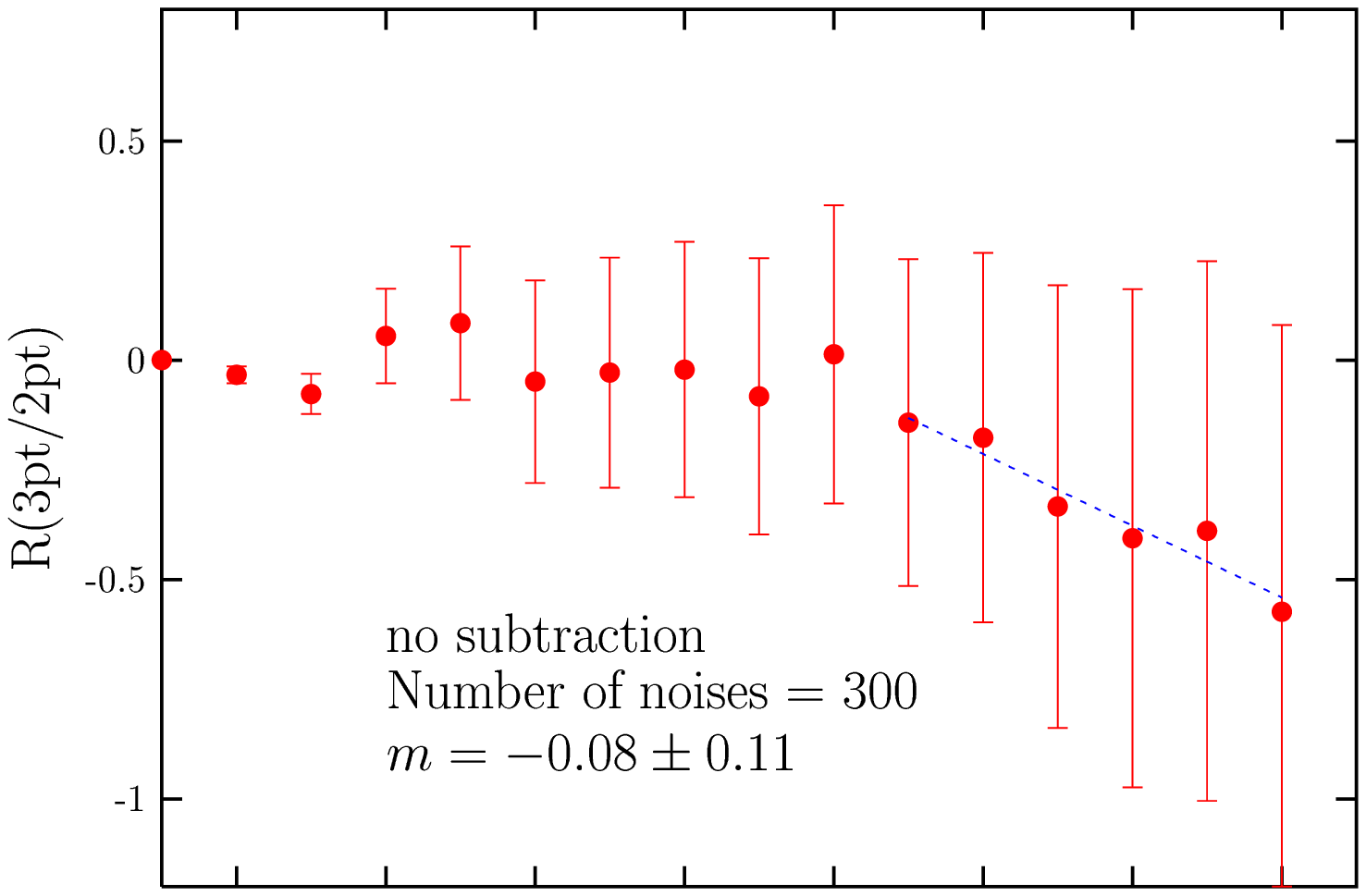}
\includegraphics{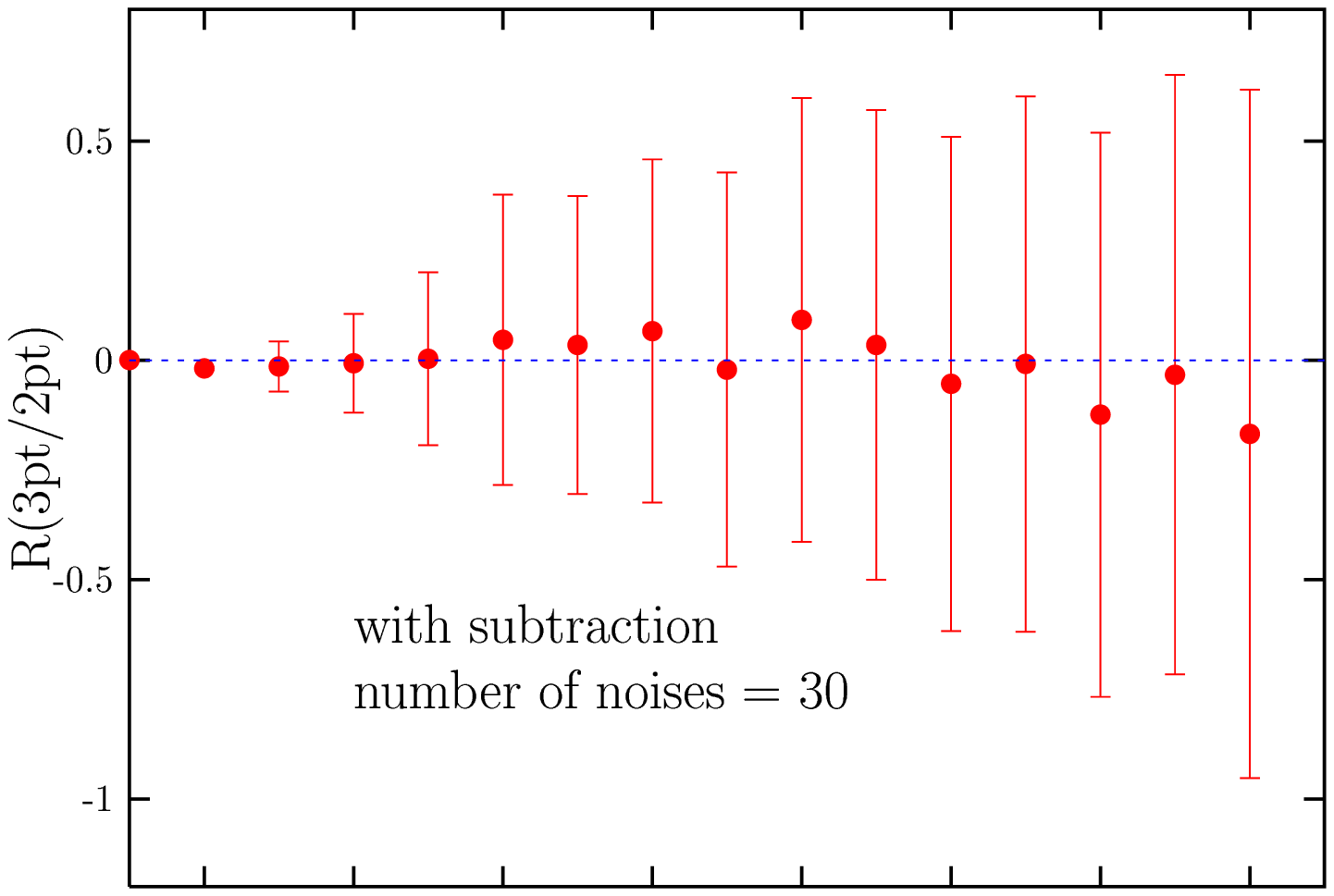}
\includegraphics{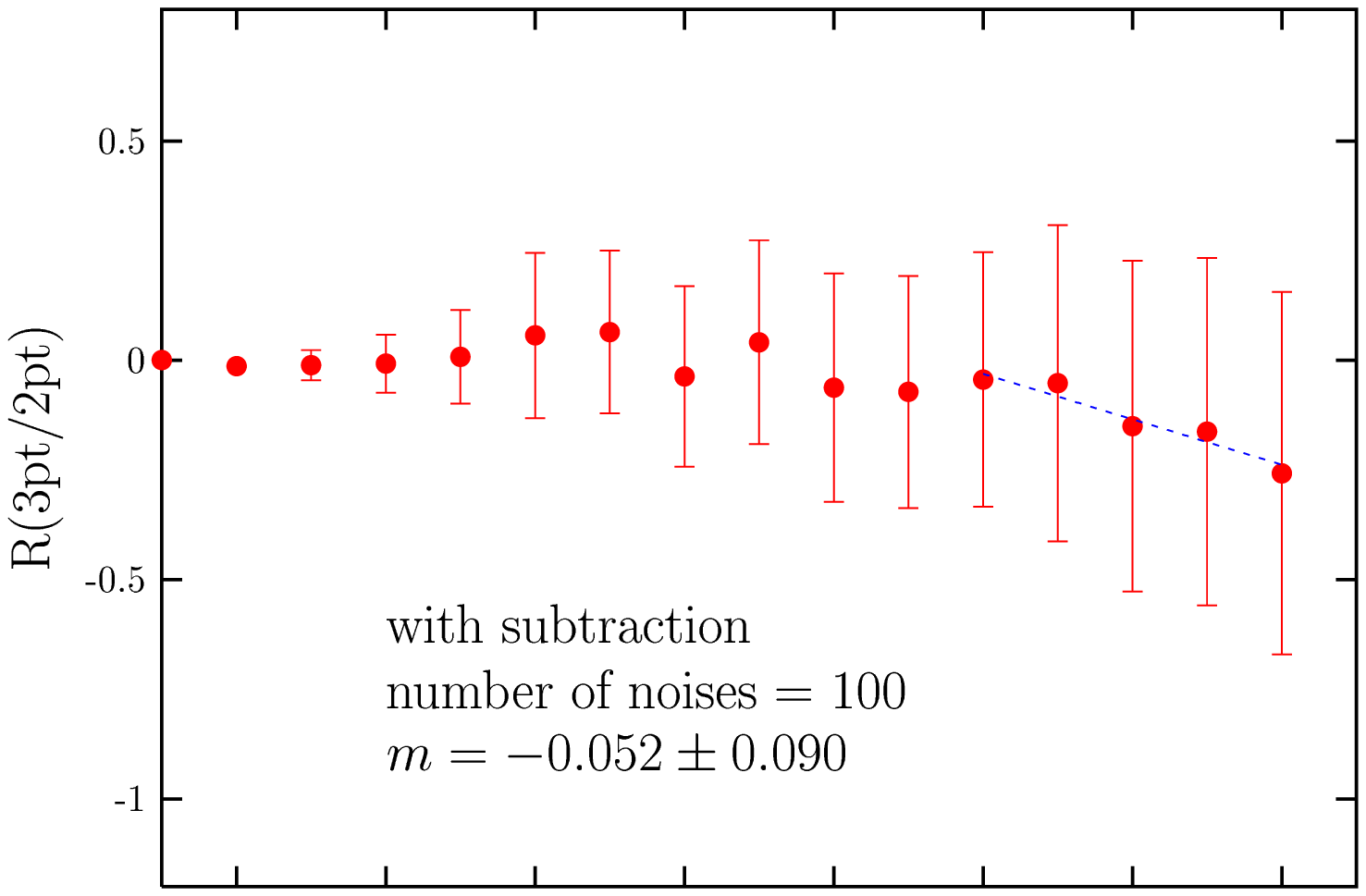}
\includegraphics{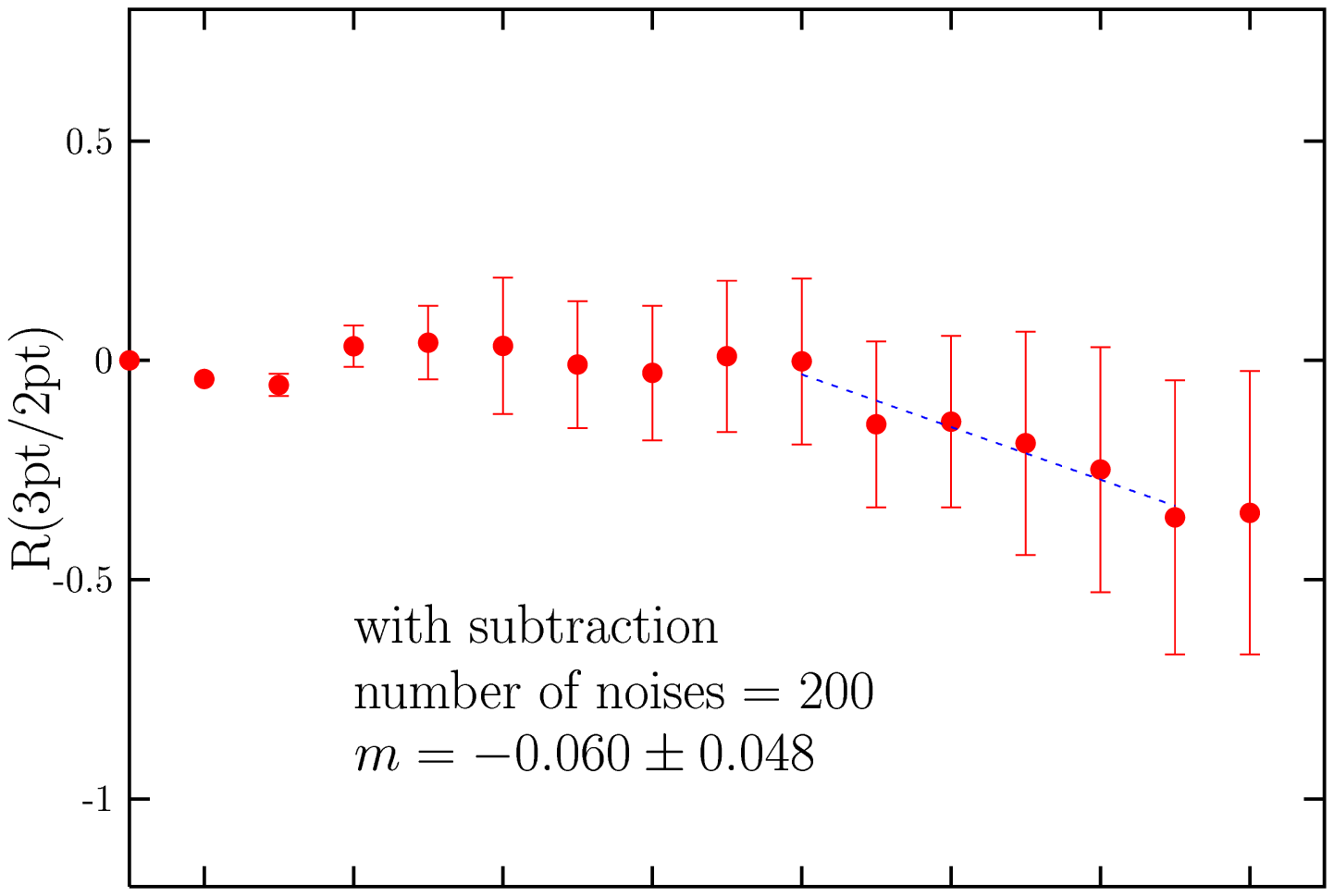}
\includegraphics{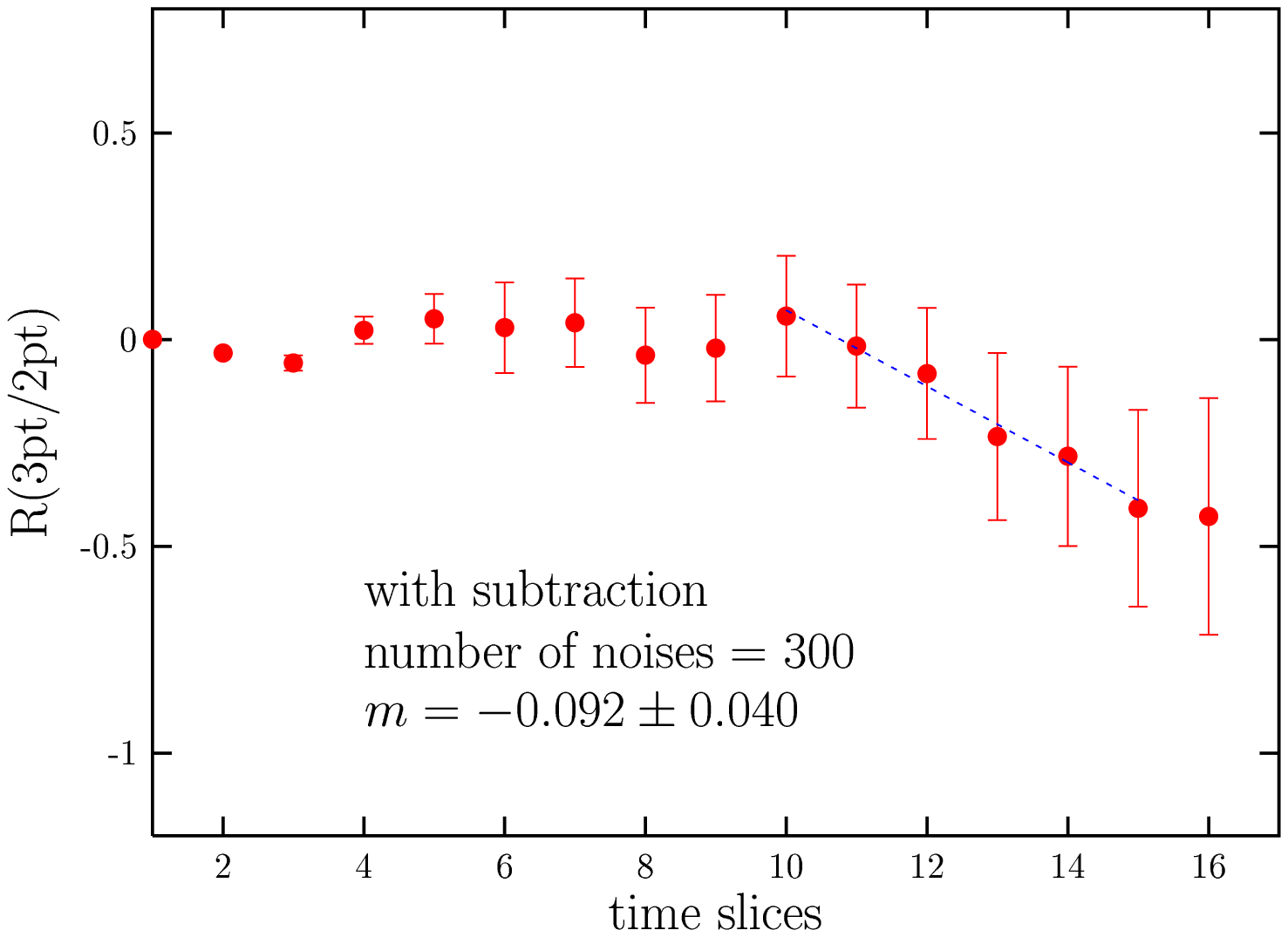}
\vskip 6.9in
\caption{Summed ratio of three to two point functions for 
different number of noises from 30 to 300. $m$ is the fitted slope
which is related to the magnetic form factor \cite{vec1}.}
\vskip -0.4in
\end{figure} 

\vspace*{-0.1in}

\section{SUMMARY}
We use noise method to extract the disconnected quark loops. As an example, we choose 
the strangeness magnetic 
form factor of the nucleon. An unbiased subtraction method is employed to reduce the variance 
in trace estimation. This study suggests that certain minimum number of $Z_{2}$ noises are 
required to extract the signal. In the case of the strangeness magnetic form factor we need 
around 300 complex $Z_{2}$ noises. Results of this study is 
consistent with our previous results \cite{vec1}. We believe, this also explains why with 60 real $Z_{2}$ noises, the work of \cite{vec2} did not see a signal even with a larger number of gauge configurations. In future we hope to carry out 
this strangeness calculation with the overlap fermion.

\end{document}